# The SIRIUS Mixed analog-digital ASIC developed for the LOFT LAD and WFM instruments

A. Cros*[ab], D. Rambaud[ab], E. Moutaye[ab], L. Ravera[ab], D. Barret[ab], P. Caïs[cd], R. Clédassou[e], P. Bodin[e], JY. Seyler[e], A. Bonzo[f], M. Feroci[gi], C. Labanti[h], Y. Evangelista[gi], Y. Favre[j]

[a]Univsité de Toulouse, UPS-OMP, IRAP, Toulouse, France;
[b]CNRS, IRAP, 9 Av. colonel Roche, BP 44346, F-31028 Toulouse cedex 4, France;
[c]Univ. [c]Bordeaux, LAB, UMR 5804, F-33270, Floirac, France;
[d]CNRS, LAB, UMR 5804, F-33270, Floirac, France;
[e]CNES, av. E.Belin Toulouse, France;
[f]Dolphin Integration, Grenoble, France;
[g]IAPS-INAF, Via del Fosso del Cavaliere 100, I-00133, Rome, Italy;
[h]IASF/Bo-INAF, Via Gobetti 101, Bologna, Italy;
[i]INFN - Istituto Nazionale di Fisica Nucleare, Sezione di Roma Tor Vergata, Via della Ricerca Scientifica, 1, I-00133 Roma, Italy;
[j]DPNC, Geneva University, Quai Ernest-Ansermet 24, CH-1211,Geneva, Switzerland.

## ABSTRACT

We report on the development and characterization of the low-noise, low power, mixed analog-digital SIRIUS ASICs for both the LAD and WFM X-ray instruments of LOFT. The ASICs we developed are reading out large area silicon drift detectors (SDD). Stringent requirements in terms of noise (ENC of 17 e- to achieve an energy resolution on the LAD of 200 eV FWHM at 6 keV) and power consumption (650 µW per channel) were basis for the ASICs design. These SIRIUS ASICs are developed to match SDD detectors characteristics: 16 channels ASICs adapted for the LAD (970 microns pitch) and 64 channels for the WFM (145 microns pitch) will be fabricated. The ASICs were developed with the 180nm mixed technology of TSMC.

## 1. INTRODUCTION

### 1.1 Context of the development of the ASIC

LOFT was proposed to ESA as the Cosmic Vision M3 mission [10], [14], and not selected. LOFT was dedicated to the study of strong field gravity and the equation of state of dense matter. LOFT consisted of two instruments: a Wide Field Monitor (WFM) and a Large Area Detector (LAD) [11]. Both were based of X-ray sensitive large area Silicon Drift Detectors (SDD) [13] to be read out by a low noise, low power, analog-digital ASIC developed under IRAP responsibility. The design was partially under contracted to Dolphin Integration under CNES funding. Stringent requirements on the noise performance were dictated by the low-energy threshold of the WFM (50 eV) and the spectral resolution of the LAD (200 eV FWHM at 6 keV and at -30°C). To match the large effective area of the LAD (10 m$^2$), about 30000 16 channels flight ASICs were needed for LAD and 700 64 channels ASICs for the WFM. The instrument design is described in detail in reference [10] and [11]. In Figure **1-1**, the two instruments are shown.

Below we describe briefly the interfaces between the SIRIUS ASIC and SDD, discuss on the performance requirements, the SIRIUS 1 and SIRIUS 2 design, the test results of the first prototype, and the first test results of SIRIUS 2. We then conclude on possible improvements to further reduce the ENC.

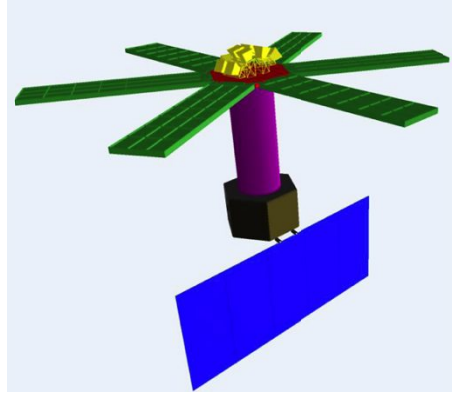

Figure 1-1: Loft payload in its original M3 proposal configuration. At the top, the 6 panels of LAD (10 m$^2$ effective area) and in the center the 6 cameras of WFM: green = LAD, yellow = WFM, red = Optical bench, purple = Structural Tower.

The design of the first ASIC prototype, called SIRIUS1, has started in April 2012 and was submitted for production end-September 2012. The first ASIC prototype was aimed to demonstrate the most critical analog performances (low-noise and low-power). A second prototype, called SIRIUS2, was manufactured with the full functionalities required by the LAD instrument. It is actually tested at IRAP for functionalities and performances.

## 1.2 Interfaces with SDD and Front End Electronics

Each LAD detector (112 anodes per side) is glued on the front side of a PCB. Sixteen detection chains are implemented in each ASIC to fit with the size and the number of anodes of the detector. Fourteen ASICs are glued on the rear side (7 per detector side). The ASIC control/test pads are bonded to the PCB and the sixteen inputs directly to the SDD anodes, to reduce as much as possible the capacitance of the bonding's and so minimize the induced noise.

Each ASIC has a SPI like interface with the Module Back End Electronics (MBEE) to receive commands, transmit data and get the main clock (Frequency 10 MHz) used for digitization. A trigger line is provided to indicate that an event has occurred. An internal register (trigger map) indicates which anode was hited. This trigger is provided to the adjacent ASICs in order to hold and then read the 7 ASICs placed on 1 side of the SDD, for noise measurements and common mode correction.

A reference voltage will be used for gain calibration purpose (internal capacitor switched from ground to anode inputs to inject a charge). This reference voltage is internally generated (five bit DAC) and level controlled by the MBEE. Finally, energy calibration will be performed with X ray sources, after integration with a detector.

Each ASIC has four channels with buffered outputs for CPA, shaper and peak and hold, which could be used for testing and calibration purpose. The power dissipation of these drivers will be reduced (close to 0 μW) during normal operations by commanding the buffers off.

## 1.3 Performance requirements

The requirements and the goal of the SIRIUS ASIC are summarized in the Table **1-1**: the LAD energy resolution is 200 eV FWHM over single-anode events at 6 keV and at end of life of the detector.

Table 1-1: Main requirements impacting the design of the analog section of SIRIUS2

| Item | Requirements/Goal | Comments |
|---|---|---|
| Energy range | 2 – 30 keV /1.5 – 80 keV | 2-80 keV expanded |
| Spectral resolution (FWHM) | <260 eV / 200 eV<br>200 eV /160 eV | @ 6 keV (end of life)<br>(for single events, 40%) |
| Energy scale | 1% / 0.8% | |
| Energy lower threshold | 200 for LAD<br>50 for WFM | 750 eV (LAD) and 350 eV (WFM) are the limits where the SDD sensitivity to X rays is decreasing |
| Dead time | < 1% / 0.5% | @ 0.7 counts/s/anode |
| Maximum count rate | 10 / 20 counts/s/anode | |
| Total dose for mission duration | 1 kRad | Low earth equatorial orbit |
| SEU rate per ASIC | LET > 60 MeV | This requirement is not a driver. (1 delatcher is implemented per 1/2 detector) |
| Bit flip mitigation | Yes | Requires triplication of registers |
| Detector type | Silicon Drift Detector (SDD) | (LAD and WFM) |
| SDD anode pitch | 970 μm (LAD) | 145 μm (WFM) |
| Number of anodes per SDD | 112 x 2 | 112 per side |
| SDD capacitance | 350fF (LAD), 81fF (WFM) | |
| SDD leaking current end of life | ~ 6.4pA (LAD), ~ 3pA (WFM) | |

## 2. SIRIUS DESIGN

Two ASICs have been developed and produced in 2 years, (see Table **2-1**) demonstrating:
- State-of-the-art electrical performances in term of noise of the Analog Front End
- High reactivity of both Dolphin Integration and IRAP to cope with the design, characterization and improvement of the analog and digital design
- The control of the interface of high performance analog blocs and digital back end in term of functionality and interference

Tests on the SIRIUS2 are in progress….

Table 2-1: Sirius Odyssey

| | State-of-the-art 2012 | SIRIUS 1 : Q1-2013 | SIRIUS 2 : Q2-2014 |
|---|---|---|---|
| Input charge range | 500 eV – 50 keV | 200 eV – 80 keV | 200 eV – 80 keV |
| ENC (at -30°C) | < 30 electrons rms | < 20 electrons | < 20 electrons |
| SDD EOL leakage current | < 2 pA | < 10 pA | < 10 pA |
| Shaping peak time | 1-10 μs programmable | 2-8 μs programmable | 2-8 μs programmable |
| ADC resolution | 9-10 bits | | 13 bits |
| Dead-time | Not communicated | 0.7 % | 0.7 % |
| Baseline restoration | Not communicated | 50 μs | 50 μs |

### 2.1 Choice of ASIC technology

Two processes have been evaluated: a space qualified 0.18 μm technology from ATMEL and a commercial 0.18 μm analog process from TSMC. Although not space qualified, and not withstanding the lack of hardened libraries, TSMC offers significant advantages: frequent prototype shuttles (Multi Project Wafers: 10 per year vs 2 per year for Atmel), rich libraries and well characterized analog models. TSMC process was chosen for the fast turnaround. This fastens the availability of prototypes, securing characterization process and significantly reducing risks within the reduced time frame. The exact technology used is TSMC foundry, 180nm process, general purpose mixed-mode, 1.8v/3.3v, variant 1P6M (1 layer for power distribution, 6 metal layer,), 4 Mx and 1 Mz metal stack configuration with MIM ($1pF/\mu m^2$), deep NWELL option.
At IRAP, we have used the facilities of AIME (Atelier Inter-universitaire de Micro/nano-Electronique) for the design (TSMC libraries, computers and tools) and Europractice IC Service, (offered by IMEC and Fraunhofer, for low-cost

ASIC prototyping, Belgium) to interface with the TSMC foundry.

The total radiation dose expected on the LOFT orbit is low, so there is no need for a hardened technology. The ASIC was designed using the TSMC rules to reduce the SEU effects and to increase the LET (Figure **2-1**): guard ring around the various functions, ring of the opposite polarity around transistors when NMOS and PMOS transistors are close, increased distance of active zone from the pads when they are connected to a pad.

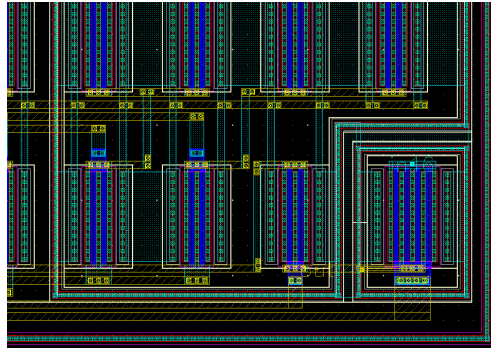

Figure 2-1: Detail of the layout showing the rules (rings) used to reduce the susceptibility to latch-up

## 2.2 Sirius 1 design

The first test chip focused only on analog performance validation.

**Channel design**

Each channel (Figure **2-2** and Figure **2-3**) includes a Charge Pulse amplifier (CPA) with a reset capability and a calibration circuit (to allow gain calibration, functional and performance testing before coupling with the detector), a CR-RC$^2$ Shaper and a Peak and Hold circuit (P&H)). Height identical analog chains are included in the ASIC, four of them at the LAD pitch and the 4 others at the WFM pitch.

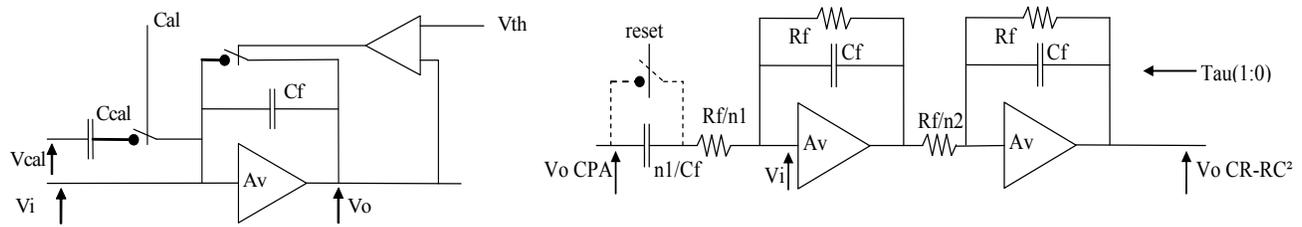

Figure 2-2: CPA and Shaper block diagram

**Charge Pulse Amplifier (CPA):**
It has been shown that resistance based leaky integrator can meet the critical specifications for LOFT project only with very large resistors practically difficult to use. So, a reset circuit has been developed to restore the base line after each event and when the baseline drift is such that a threshold is reach. A capacitor is used to inject charges for calibration purpose. CPA characteristics are: minimum gain 60dB, max power consumption 250µW, feedback capacitor 75fF.

**Shaper**
Pulse shaper is a single ended simple CR-RC$^2$ semi Gaussian filter. Inverting configuration allows for a simple common mode voltage buffer with low drive requirements. Offset is less than 5 mV per amplifier. The time constant is set by a switched network capacitors and selection logic. This allows maintaining the input impedance (at high frequency) identical regardless τ. Four peaking time can be selected (2, 4, 6 and 8 µs). Shaper characteristics are: gain 8 (first stage) and gain 5 (second stage), max power consumption 200µW.

**Peak and Hold**
Various designs have been studied and we retain a peak detector and switched capacitor based hold amplifier. A switched capacitor hold suffers from switch injection errors, but careful design keep the remaining errors low enough. Peak detection is performed by voltage derivation and zero detection, and for best performance, separate Peak detection and Hold functions are used. The feedback capacitor is selected according τ value to optimize the detection. A 5 pF hold capacitance, Ch, provides good noise and retention performances at reasonable power consumption (target: 150 µW worst case).

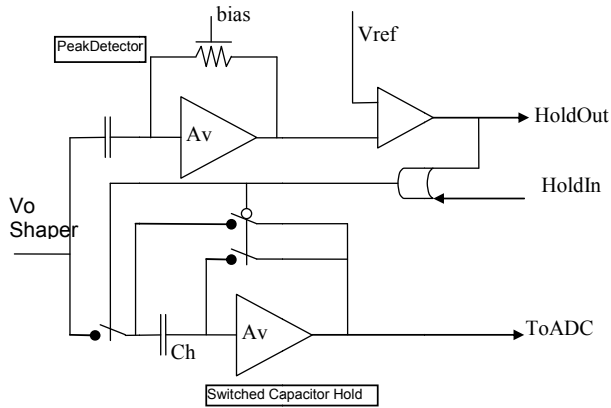
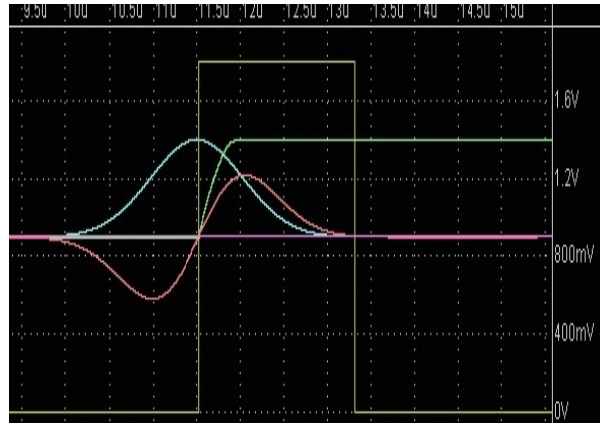

Figure 2-3: Peak & Hold block diagram (left) and hold signal due to an event (right). The blue curve is the output of the shaper; the pink one is its derivation (when crossing the zero, the hold signal, yellow, is generated). The green line is the output of the shaper going to the ADC.

## ENC

ENC is the value of charge that, injected across the detector capacitance by a delta-like pulse, produces at the output of the shaping amplifier a signal whose amplitude equals the output r.m.s. noise (Veff).

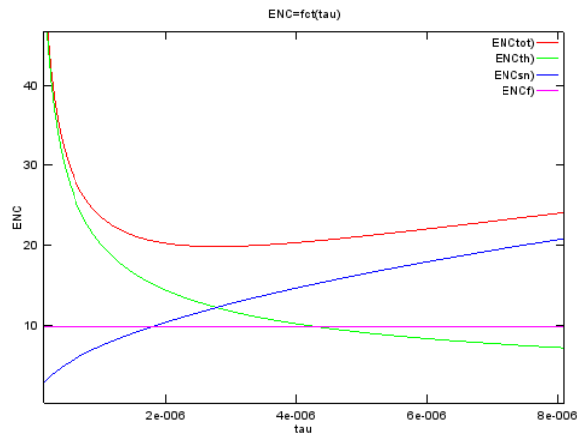

Figure 2-4: Simulation showing the noise performance of the 1$^{st}$ IRAP/Dolphin ASIC prototype, as a function of τ (switched reset, 10pA SDD leakage current, τ in s). $ENC_{tot}$ (red) is the square root of the sum of the square of the various ENC ($ENC_{th}$: CPA thermal noise, green; $ENC_{sn}$: shot noise due to detector, blue; $ENC_f$: 1/f noise, purple)

ENC simulations (Figure **2-4**) have shown that the ENC of the integrated Front End Electronics (SDD + ASIC) specified for the first prototype (20e-) can be achieved with a SDD end of life leakage current of 10pA, at -30°C, with a tau of 3 µs.

### 2.3 Sirius 2 design

The second prototype (SIRIUS 2), jointly developed by IRAP and Dolphin Integration, is a version of the ASIC implementing all the functionalities required for LAD-LOFT: various control capabilities and serial interface (SPI) with the Front End Electronics, LAD size and pitch, 16 channels, PGA and 10/12/14 bit ADC, threshold control (16 x 8 bit DACs), calibration DAC (5 bit), reference band gap and internal test feature. The topology of the ASIC was also changed: every detector inputs placed at one side of the chip and the control on the opposite side to comply with size of the SDD and to reduce the coupling between digital section and analog blocks. The design of CPA, shaper and S&H was kept identical.

The second prototype will be used to perform radiation tests, aimed at qualifying the ASIC design and technology for Total Integrated Dose (TID) and Latch-up.

**SIRIUS2 block diagram**

Reference band gap, multiplexers, 8 bit DACs, 5 bit DAC and control logic were developed by IRAP. Dolphin Integration was in charge of the integration of the 16 analog chains, the verification of the design rules, the integration at the top level, mixing analog and digital modules, and of the adaptation of a Dolphin IP ADC for SIRIUS2 design.

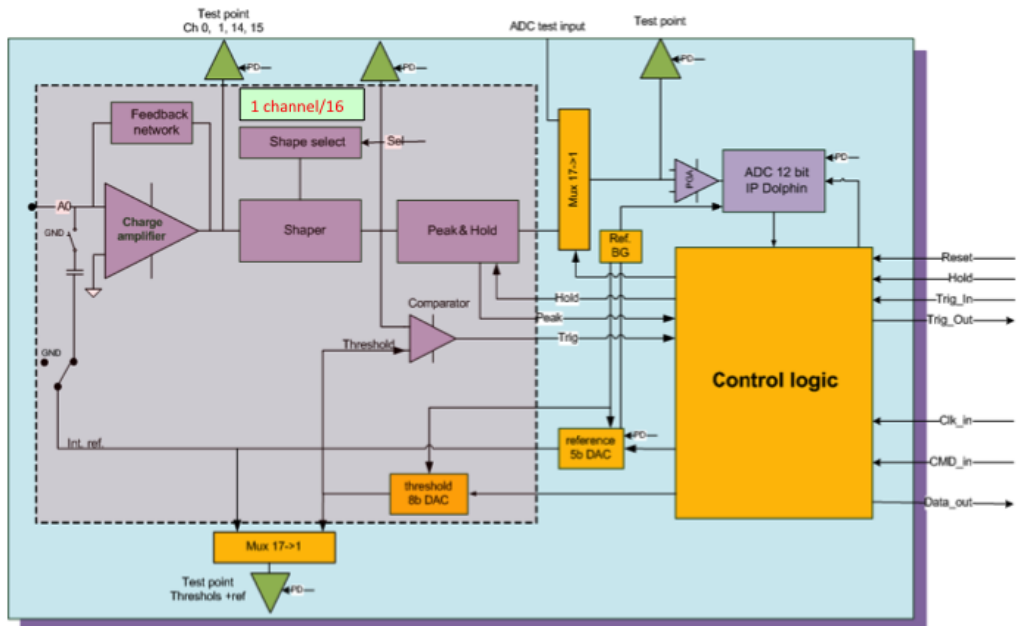

Figure 2-5: Functional block diagram of the SIRIUS2 ASIC designed by IRAP and Dolphin. Orange color indicates the functions designed by IRAP. Only 1 analog channel among 16 is shown.

**Functionalities and control**

The digital electronics is providing a serial SPI like interface for internal control and data reading. Functionalities and extra pads were added for testing purpose. Each analog function can be tested separately.

*Configuration commands*: it is possible to select the value of the peaking time (4 values), to disable any channel individually in case of noisy SDD anode, to define the 16 detection thresholds, to choose the level of charge injection, to adjust and optimize the reference band gap versus temperature and adjust the reference voltages, to define the operating mode of the ADC, to select the channels tested, to generate a calibration sequence

*Operation commands*: commands to read the trigger map, to digitize the P&H output, to read the 16 channels, to reset the triggers after an event.

*Test commands*: these commands are mainly used for testing/calibration purpose during the development phase but could also be used during operations: switch ON/OFF of various sections of the ASIC (Bias, CPAs, Shapers, Peak & Hold, ADC, DACs, Buffers, Calibration DAC, ADC buffer) in order to reduce as much as possible the power consumption during the operations; configuration of multiplexers to address various internal signal and send them to some test pads; reading of the internal reset counter.

## 3. TEST RESULTS: FUNCTIONALITIES AND PERFORMANCES

### 3.1 SIRIUS1 Layout

The 4 upper channels are the LAD channels (970 µm pitch) and the 4 lower the WFM one (145 µm pitch) (Figure **3-1**). Twenty ASICs were encapsulated for standalone test purpose and 8 were glued directly on a test board to be integrated with a SDD detector (Figure **3-6**). The pads connected to the detector anodes are on the left side and the power distribution, the digital control and buffered outputs on the 3 other sides.

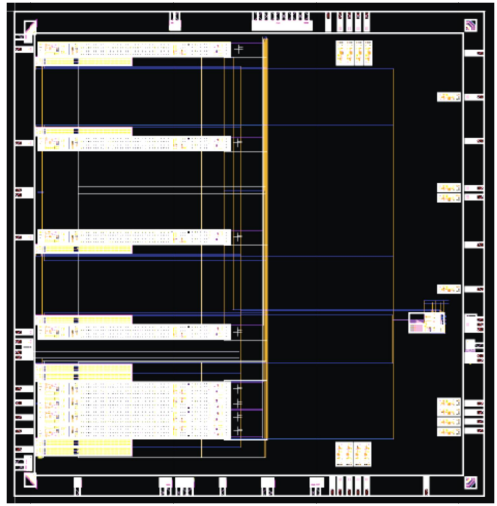

Figure 3-1: First prototype layout of SIRIUS1 (Silicon proof): 5x5 mm$^2$

## 3.2 Test setup and tools

For testing, the encapsulated ASIC is installed in a support (Figure **3-2**) on a test board developed by LAB (University of Bordeaux, CNRS). Inputs of the ASIC (normally connected to the SDD anodes) are not bonded inside the package to not add capacitive load at input of the CPA. The only capacitance is due to the input PAD and its ESD protection (around 100fF). By this way it is possible to measure the intrinsic performance of the ASIC alone. Perturbations due to the SDD detector are removed (SDD capacitance, bonding capacitance, SDD noise due to leaking current).

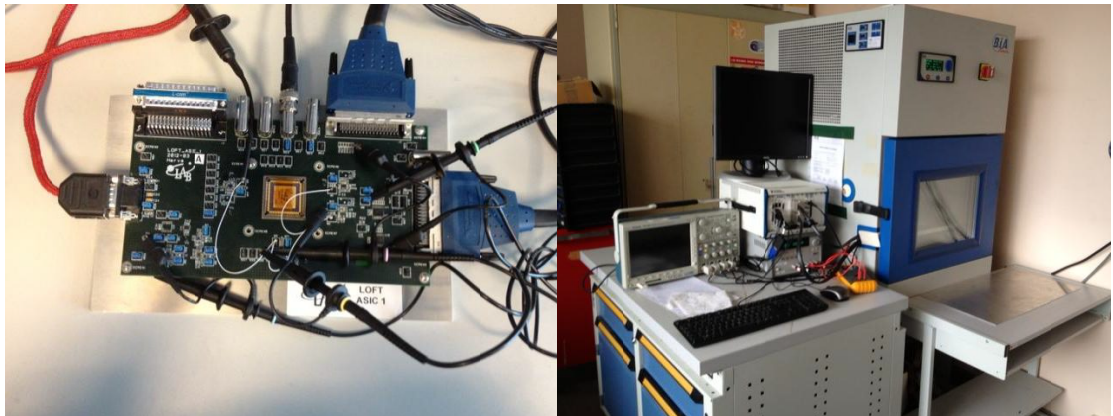

Figure 3-2: Test board (left) and test setup for thermal tests (right)

The tests tools use modules installed in a PXI Express chassis: controller with up to 24 bit Digitizers, Virtex 5 FPGA and Multi bank multiplexer.

## 3.3 Sirius 1 test results

The SIRIUS1 ASIC was tested alone at IRAP, Toulouse, and with the SDD detector at INAF/IASF, Bologna, Italy.

**Power consumption**
The ASIC was powered by an external power supply that provides +5V for the test board electronics. Linear regulators provide the 3,3V to the digital I/0's of the ASIC. A clean low noise power supply provides the 1,8V to the internal analog and core digital electronics of the ASIC. It is possible to power or switch off independently the bias circuits, the 8

CPA's, the 8 Sampler's, the 8 Sample & Hold's and the 12 buffers. We measured (Figure **3-3**) at each temperature, the power consumed by the 8 analog chains (buffers off). The power is lower than expected based on simulations, and within specifications.

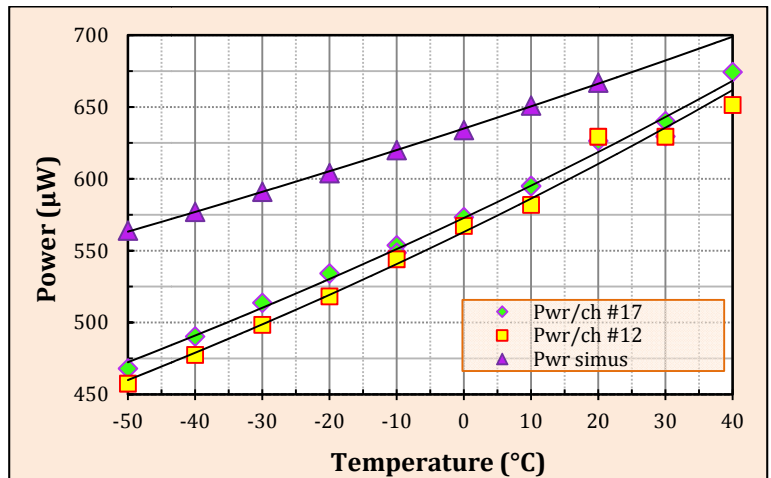

Figure 3-3: Measured power consumption of 2 ASIC versus temperature (green lozenge and yellow square) compared to post layout simulations (purple triangle)

**Gain and Linearity**

The built-in injection capacitor is used to inject various charges at the input of the CPA. For each temperature, a linear fit of Vout is calculated (a and b parameters) and the quality of the fit is estimated ($R^2$ parameter) (see Figure **3-4**).

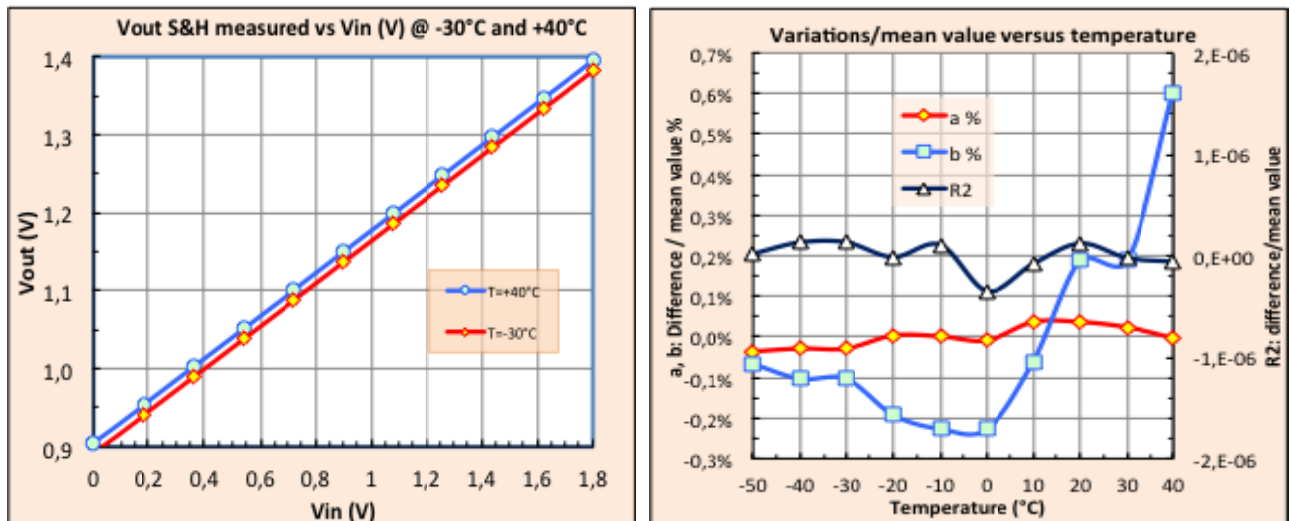

Figure 3-4: Gain measured @ 40°C and @ -30°C (left) and variations of the gain, offset and $R^2$ coefficient versus temperature (right). Vin is the injection level: 0V for 0 eV and 1,8V for 80 keV. R2 is very close to 1 and the variation is less than 0,7 E-6.

We can conclude from the $R^2$ factor that the fit by a linear regression is very good. The measured variations of gain (less than +/- 0,04%) and of the offset (less than +0,6% / -0,3%) in the full temperature range (-50°C to +40°C) show that the gain and the offset are stable versus temperature.

**Noise (ENC)**

We used the ASIC built-in capacitor to inject a charge at the input of the CPA. For each injection 1000 measures are made of the hold voltage using a 24-bit ADC. The mean value is calculated to minimize the noise due to the test equipment. One thousand identical charges are injected for each injection level. The mean value of the output voltage is computed and the sigma of the distribution of the 1000 measures. The value of noise is computed from this sigma value. It is not varying much according to the injected level. We repeated the same measures at various temperatures (see Figure **3-5**). The ENC is decreasing with the temperature to reach 16.5 e$^-$ below -30°C.

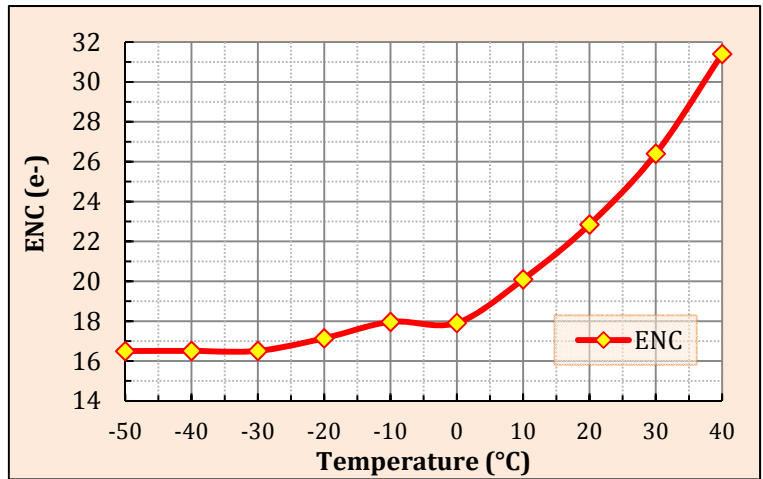

Figure 3-5: Variation of the mean noise of the ASIC alone, versus temperature, to be compared with the post-layout simulations: 27,3 e- @ +25°C and 13 e- @ -30°C

### 3.4 Performances after Integration with a SDD detector

**Test setup and test board**

The SDD with LAD pitch on one side and WFM pitch on the other (4" FBK detector) is mounted on a board developed by DPNC (University of Geneva). It is enclosed in a metallic box. On the picture (Figure **3-6**), four ASIC are visible on the bottom and two on the top. Their inputs are directly bonded to the detector anodes.

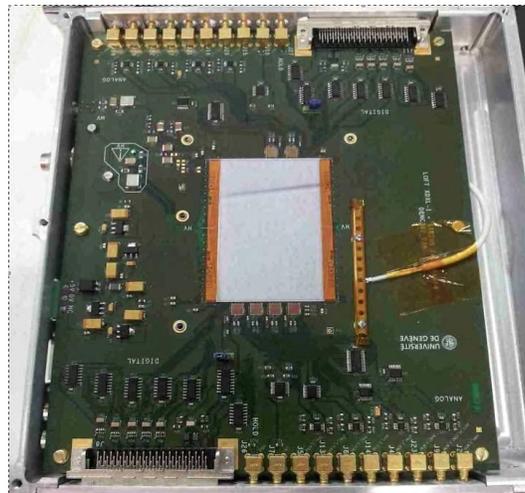

Figure 3-6: Test board. The SDD Detector is in the middle. On the bottom of the detector (LAD side of the SDD), 4 SIRIUS1 ASICs are glued and bonded directly to the anodes of the SDD, and on the top of the detector (WFM side), 2 other SIRIUS1 ASICs are also glued and bonded.

**Spectra from (buffered) shaper output**

With a very simple setup a spectra is collected from one channel of the SIRIUS1 ASIC connected to an anode of the LAD side of the SDD. The shaped signal (2 µs peaking time) of ASIC #4/ Channel 1 was directly connected to a multichannel analyzer. Figure **3-7** below shows the spectrum acquired.

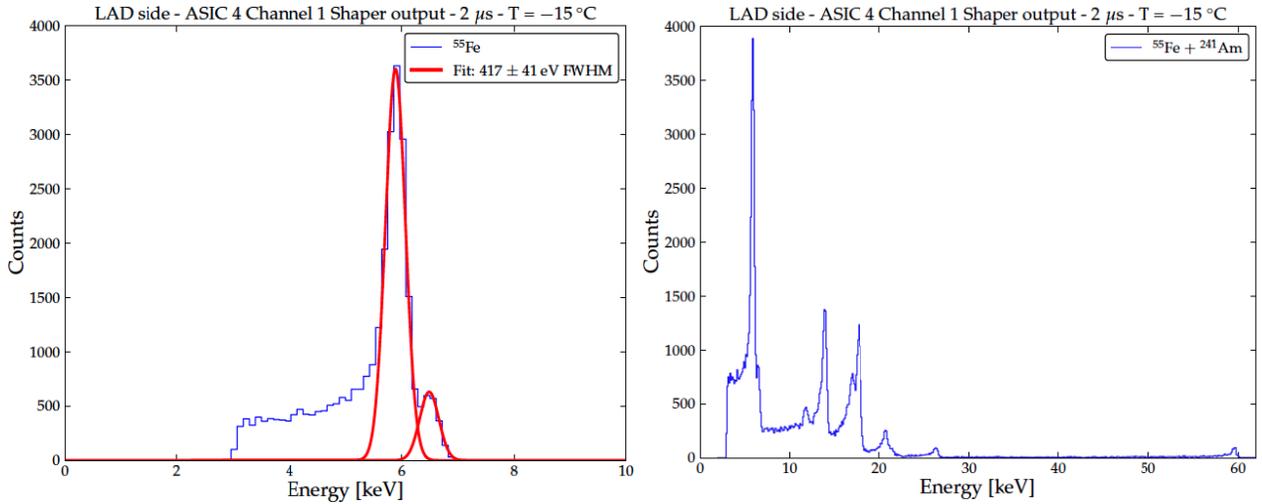
Figure 3-7 Left plot: Fe-55 alone - Right plot: spectra with both Am-241 and Fe-55 sources.

The resolution of the peaks, (see Fe-55 spectra) suffers for the charge sharing effect between two or three adjacent anodes that cannot be taken into account with this simple setup. The raw spectrum of a single channel acquired with the shaper signal includes not only the complete charge collection for one event, by the connected anode (single-anode events), but also the partial charge due to events impinging near the adjacent anodes. This is reflected in the tail on the left side of the peaks.

### 3.5 Sirius 2 test and performances

**SIRIUS2 Layout**

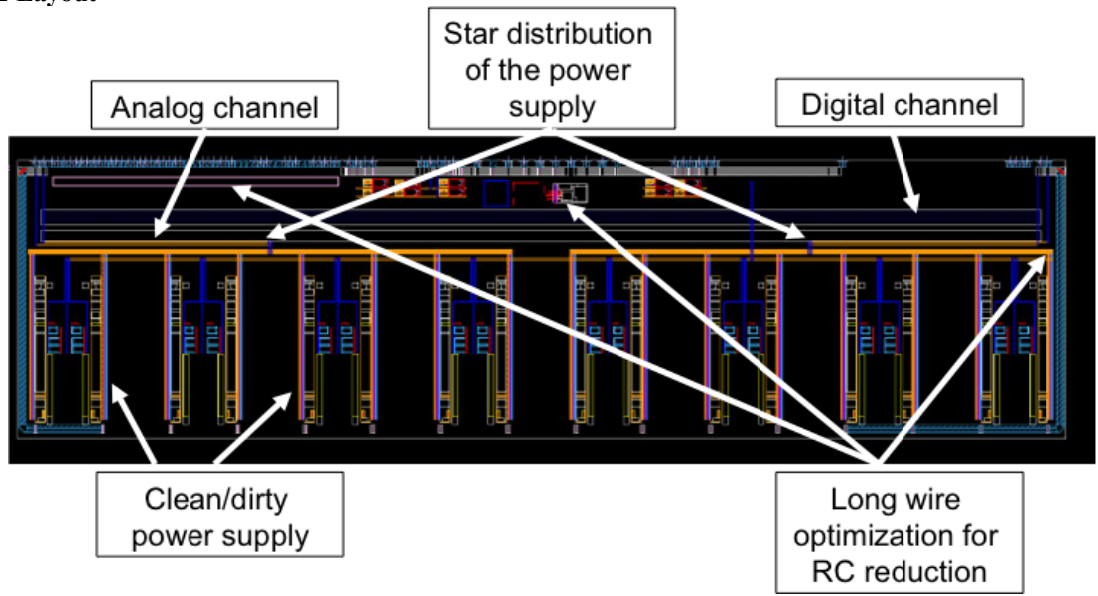

Figure 3-8: SIRIUS 2 layout. The 16 analog channels are visible at the bottom with the 16 input pads. The upper section is dedicated to control electronics, multiplexers, DACs, PGA, ADC, reference band gap, I/O and tests pads.

**Test equipment:**
A High Density of Integration PCB (Figure **3-9**) has been developed by LAB to support the test of SIRIUS2 ASIC: more than 50 bonding are necessary to feed the inputs/outputs; the 16 CPA input pads are left unconnected. Low noise power supplies regulators for the ASIC and the associated filtering, are installed on this board. It is powered with +5V. A temperature sensor placed close to the ASIC is used to measure the temperature of the ASIC. It is readable through a SPI like interface. The connection with the control PC is made through an Opal Kelly board plugged directly on the test board (acquisition/transmission of analog and digital signals + SPI).

Automatic test sequences have been built, including data acquisition from voltmeter, display recovery from scope, plot generation and data analysis.

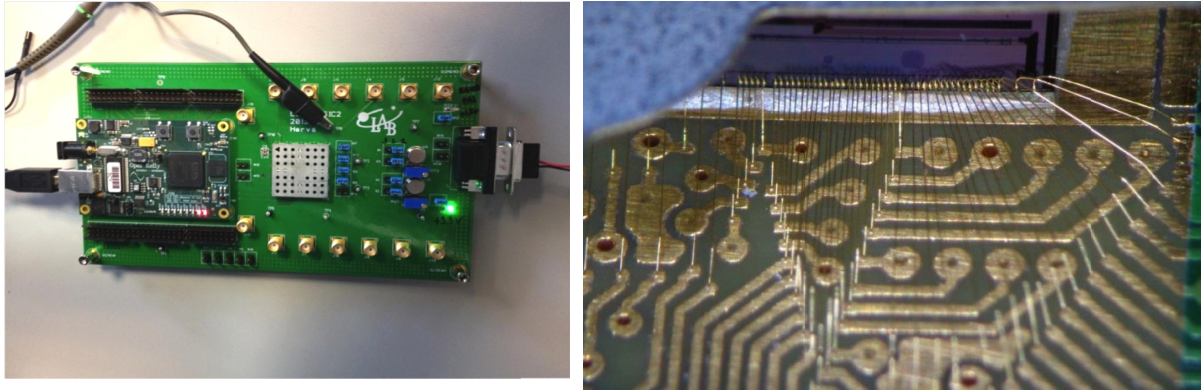

Figure 3-9: ASIC board with Opal Kelly daughter board installed (left) and ASIC bonding (right).

**Functional preliminary tests**

Preliminary results show that the digital control of the various functions is operating as expected: peaking time selection, reset, Reference Band gap/VCM, 8 and 5 bit DACs, Multiplexers, charge injection circuit and associated timing circuits. Reading of the 16 channels after digitization is also operating as expected.

**Performance**

Test results not available today.

## 4. POSSIBLE IMPROVEMENT FOR FUTURE DEVELOPMENT

### 4.1 Theoretical and simulation studies

We have performed some theoretical studies and various simulations to found how to improve the performance (especially to lower the ENC) of the SIRIUS2 ASIC. The theoretical studies are based on reference [1]. We have considered the 4 various noises impacting the ENC at the output of the shaper and tried to reduce each of them. To get the lower ENC, we already choose to not use a resistor in parallel with the feedback capacitor (see 2.2). The remaining noises are the thermal noise ($ENC_d$), the flicker noise ($ENC_f$) and the detector noise ($ENC_o$). This last noise is not considered now (not related to the ASIC).

Based on ref. [1], the expressions of the various ENC are:

ENC due to **CPA thermal noise**: $ENC_d^2 = \dfrac{8kTC_t^2 n(n!)^2 e^{2n} B\left(\frac{3}{2}, n-\frac{1}{2}\right)}{12 g_m q^2 \pi \tau_s n^{2n}}$  Equation 1

ENC due to **1/f noise**: $ENC_f^2 = \dfrac{K_f C_t^2}{2q^2 C_{ox}^2 WL} \dfrac{(n!^2) e^{2n} B(1,n)}{n^{2n}}$  Equation 2

ENC due to **detector leaking current (shot noise)**: $ENC_o^2 = \dfrac{I_0 \tau_s}{2q\pi n} \dfrac{(n!^2) e^{2n} B\left(\frac{1}{2}, n+\frac{1}{2}\right)}{n^{2n}}$  Equation 3 n is the shaper order (CR-RC$^n$), T the temperature (°K), k the Boltzmann constant, q the charge of e$^-$; $g_m$, $K_f$, $C_{ox}$ are process parameters, W and L characteristics of the CPA input transistor, B(x,y) the Beta function of x and y, $\tau_s$ the peaking time of the shaper, $I_0$ the leaking current of the detector and $C_t$ the total capacitor. In reference [1], $C_t$ is expressed as: $C_t = C_f + C_{GS} + C_{GD} + C_{det}$ ($C_f$ is the CPA feedback capacitor, $C_{GS}$ and $C_{GD}$ are respectively grid-source and grid-drain capacitors of the input transistor, and $C_{det}$ the detector capacitor at the input of the CPA.

We have computed the theoretical total ENC due to the ASIC using equations (1) and (2) and found discrepancy with simulation results. To get similar result, we have to assume a much larger capacitor (around 2100 fF) at the input of the CPA than the detector capacitor (350 fF). We consider that this extra capacitor is due to the Miller effect [12] in the CPA amplifier applied at the feedback capacitor. We have simulated the input impedance of the CPA and found a value of around 2.1pF in the interesting frequency range. To confirm our assumptions we simulate also the open loop gain of the CPA in the useful bandwidth and found a value of 31,5. If we compute the Miller capacitance at the input of the CPA, we found 75fF x 31.5 = 2.36 fF, which is really comparable to the previous value.

Based on the theoretical equations 1 and 2 we also computed the ENC of the ASIC versus tau for different values of the input capacitor of the CPA without taking into account the Miller capacitor effect. We compared these plots with the simulation of the current design. Figure **4-1** shows that, to achieve the same ENC, the input capacitance should be 2130fF. We can conclude that it is the value of the Miller capacitance.

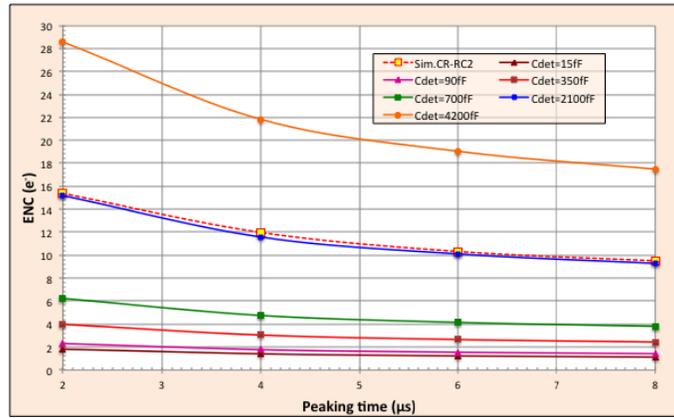

Figure 4-1: Theoretical values of the $ENC_{ASIC}$ of the current design versus peaking time for various values of the CPA input capacitor compared with the simulations (red doted line, square) of the actual design with no input capacitance (Cf=75fF), and $K_f$ = 1.77E-32 $C^2/cm^2$, CR-RC$^2$).

So, we can conclude that the Miller capacitance, $C_M$, should be added to the 4 other ones to determine the $C_t$ value used in equation 1 and equation 2:

$C_t = C_f + C_{GS} + C_{GD} + C_M + C_{det}$     with $C_f$ = 75fF, $C_{GS}$ = 183fF, $C_{GD}$ = 15.2fF, $C_M$ = 2100fF, $C_{det}$ = 350fF

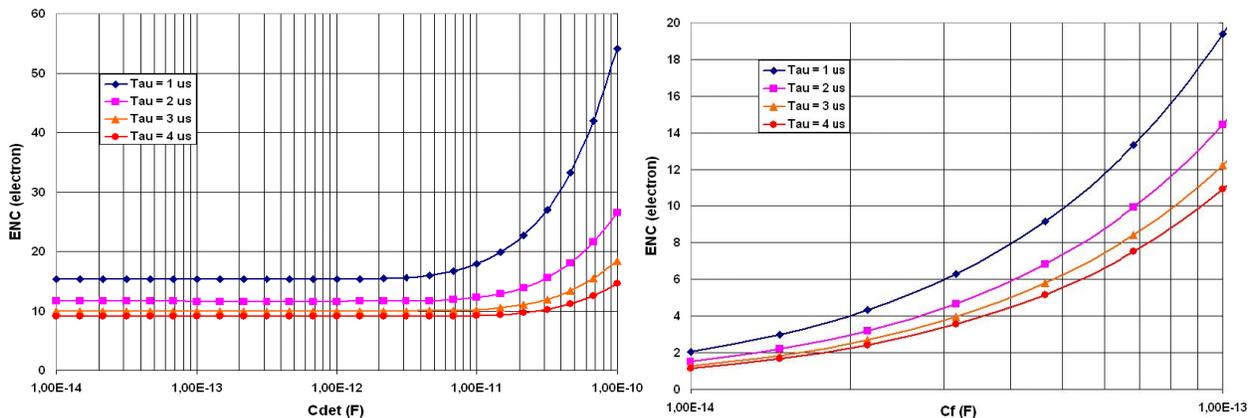

Figure 4-2: Left: Simulations of the ENC versus the entrance capacitance of the CPA for various values of tau. Input capacitance has nearly no effect up to a few pF, until it becomes in the range of the Miller capacitance at the input of the CPA (few pF). Right: ENC versus CPA feedback capacitor for low values of $C_f$ (T=-30°C)

Both ENC due to thermal noise and ENC due to 1/f noise are proportional to $C_t$. As $C_M$ is the larger capacitance in the expression of $C_t$, a way to reduce the ENC due to the ASIC is to reduce the Miller capacitance by decreasing the value of the feedback capacitance. The results of the simulations are presented in Figure **4-2**.

Finally we checked the ENC of the ASIC in the actual configuration (Cf=75fF) compared with a lower value (Cf=28fF) in the operating temperature range and for various values of the peaking time (Figure **4-3**). The very low ENC that we obtain is at the limit of the silicon technology and so shall be considered with a great care. However a significant improvement is achieved.

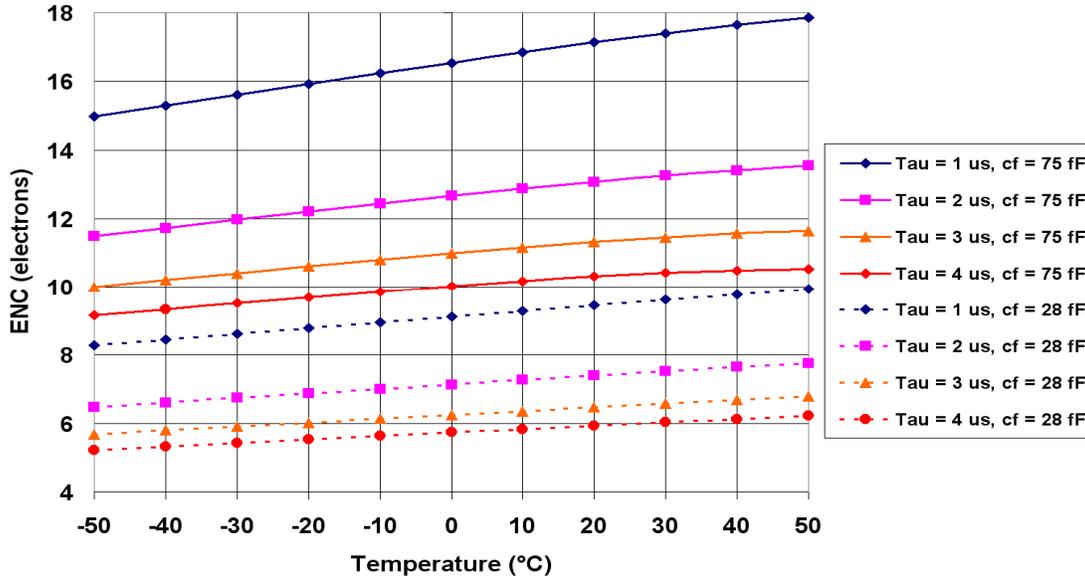

Figure 4-3: Simulations of the ENC of the ASIC versus temperature, for feedback capacitor 75fF (continuous lines) and 28fF (doted lines) for 4 values of the shaper peaking time (2, 4, 6, 8μs).

### 4.2 Consequences for LOFT

We have demonstrated that we can achieve the ENC requirement, which is critical for LOFT performances. Other important consequence for the design of the LOFT Front End Electronics is that the detector capacitance (350fF for LAD) has not a big effect on the ENC, neither the stray capacitance due to the bonding (70fF), nor ESD protection circuit placed at the input of the CPA (100 fF). The larger effect is due to the Miller capacitance.

Other possibilities for reducing the ENC have been considered (optimization of W and L, third order shaper, $CR^2$-$RC^2$ shaper). Acting on the feedback capacitor looks the easier and efficient way to reduce the ASIC ENC but these other solutions can be considered, particularly the third order shaper, because it is filtering the 3 noise sources and so, is the only way to reduce the ENC due to the detector shot noise.

## ACKNOWLEDGEMENTS

The SIRIUS2 ASIC has been developed under CNES contract in the frame of studies for LOFT. IRAP-CNRS has provided the technical support, the expertise in the X-ray detectors and associated electronics, part of the design and the management of the project. DOLPHIN Integration has provided the expertise in the TSMC technology for the analog design. Thanks to the various teams for the high motivation to achieve a high performance design, starting from crash, and in a very short delay. A special thank to Pierre Bodin for the strong encouragement and support at the beginning of the project.